\documentclass[12pt]{article}
\def\lae{\;^{<}_{\sim} \;} \def\gae{\; ^{>}_{\sim} \;}

\title{\textbf{Right-Handed Sneutrino Curvaton and non-Gaussianity}}
{\author{\\[1cm]
{\sc \large Chia-Min Lin$^{1,\dagger}$ and Kingman Cheung$^{1,2,3,\star}$}\\
{\sl\small $^1$Department of Physics, National Tsing Hua University, Hsinchu, Taiwan 300 }\\
{\sl \small $^2$Division of Quantum Phases \& Devices, School of Physics,}\\
{\sl\small Konkuk University, Seoul 143-701, Republic of Korea} \\
{\sl\small $^3$Physics Division, National Center for Theoretical Sciences,
Hsinchu 300, Taiwan}
}}

\usepackage[margin=2cm]{geometry}
\usepackage{graphicx,color}
\usepackage{subfigure}
\begin{document}
\maketitle
\begin{abstract}
  In this paper, we explore the parameter space for a Right-Handed
  (RH) sneutrino curvaton that can generate large non-Gaussianity
  without assuming any particular inflation sector. The mass of the RH
  sneutrino is suggested from a discussion on the initial condition of
  the curvaton field. It is shown that  a small Yukawa
  coupling is generally required for a successful RH sneutrino
  curvaton. However, the Yukawa coupling can be larger if we consider
  the braneworld scenario. Some general discussion about the spectral
  index in curvaton scenario is also provided.
\end{abstract}
\footnoterule{\small $^\dagger$cmlin@phys.nthu.edu.tw, $^\star$cheung@phys.nthu.edu.tw}
\section{Introduction}

As we can see from observation, the Cosmic Microwave Background (CMB)
is very isotropic but with small ($\sim O(10^{-5})$) temperature
fluctuations. The advantage of single-field slow roll inflation models
is that an inflaton field can produce inflation and also provide
primordial density perturbation as the seeds of structure formation
and CMB temperature fluctuation from the quantum fluctuation of the
inflaton field during inflation. However, the model of inflation is
highly restricted by the requirement of the right amount of primordial
density (curvature) perturbations. There is one way to liberate the
model of inflation if the job of creating curvature perturbation is
done by another field which is called curvaton \cite{Lyth:2001nq,
  Moroi:2001ct, Lyth:2002my, Linde:1996gt}. If the curvaton field is
light (smaller than the Hubble parameter), it can produce an almost
scale invariant quantum fluctuation during inflation. However, by
definition the energy density of our universe is
dominated by the inflaton field during inflation,
therefore the curvaton cannot produce
curvature perturbation during inflation. The job must be done after
inflation. If the curvaton decays after inflation, during the
oscillation of the curvaton field (described as nonrelativistic
matter), the universe will be dominated by radiation after inflaton
decay and the relative energy density of the curvaton is growing and
can produce right amount of curvature perturbation. Here we can
qualitatively know that the decay rate of a successful curvaton must
be low.

We can distinguish between the cases that curvature perturbation is
coming from inflaton or curvaton by investigating the non-Gaussianity
of CMB (for a review of non-Gaussianity see
Ref. \cite{Bartolo:2004if}). The non-Gaussianity from single-field
slow roll inflation is very small \cite{Gangui:1993tt, Gangui:1994yr,
  Wang:1999vf, Gangui:1999vg} but it can be
large from the curvaton scenario
\cite{Bartolo:2003jx, Malik:2006pm, Sasaki:2006kq}. Conventionally,
non-Gaussianity in curvaton scenario can be described by the
non-linearity parameter $f_{NL}$, which takes the form
\begin{equation}
\zeta=\zeta_g+\frac{3}{5}f_{NL}\zeta^2_g+\cdots,
\end{equation}
where $\zeta$ is the curvature perturbation in the uniform density slice and $\zeta_g$ denotes the Gaussian part of $\zeta$. Currently the upper bound of $f_{NL}$ is roughly given by ($2-\sigma$ range) \cite{Yadav:2007yy, Komatsu:2008hk, Curto:2009pv}
\begin{equation}
f_{NL} \lae 100.
\end{equation}
In the near future, the Planck satellite \cite{:2006uk} will reduce
the upper bound to $f_{NL} \lae 5$ if non-Gaussianity is not
detected. Therefore, we will consider
\begin{equation}
10 \lae f_{NL} \lae 100,
\end{equation}
which can be tested in the near future. We refer this range as large non-Gaussianity.

The possibility of using right-handed (RH) sneutrino as a curvaton was
considered in \cite{McDonald:2003xq, McDonald:2004by, Moroi:2002vx,
  Postma:2002et, Mazumdar:2004qv}. However, the parameter space for
generating large non-Gaussianity was not explored. A specific
application of RH sneutrino to D-term hybrid inflation and
non-Gaussianity was investigated in \cite{Lin:2009yn}. In this paper,
we explore the parameter space in more generally settings without assuming
any particular inflation model.

The paper is organized as follows. In Sec.~\ref{1}, we present the
formalism and describe non-Gaussianity generated in our RH sneutrino
curvaton scenario. In Sec.~\ref{2}, the initial condition of the curvaton field
is discussed. This may suggest the Right-Handed sneutrino mass. In
Sec.~\ref{3}, we discuss the spectral index in the curvaton
scenario. In Sec.~\ref{4}, we consider the curvaton on a brane, and
show that the Yukawa coupling can be larger in this case. Sec.~\ref{5}
is our conclusion.

\section{RH Sneutrino as a Curvaton}
\label{1}

The superpotential of the mass eigenstate of the RH neutrino, $\Phi$, is
given by
\begin{equation}
W_\nu=\lambda_\nu\Phi H_u L+\frac{m \Phi^2}{2},
\end{equation}
where $\Phi$ is the RH neutrino superfield, $H_u$ and $L$ are the
MSSM Higgs and lepton doublet superfields, and $m$ is the RH
neutrino mass. This gives the potential of right-handed sneutrino
$\sigma$ as follows
\begin{equation}
V(\sigma)=\frac{1}{2}m^2\sigma^2.
\end{equation}
The decay rate for RH sneutrino is
\begin{equation}
\Gamma=\frac{\lambda^2_\nu}{4\pi}m.
\end{equation}

In the following we will consider the case where the RH sneutrino is the
curvaton. During inflation we require $m \ll H$ in order to have
$\sigma$ slow-rolling, which means the field value can be taken as a
constant during inflation.

The amplitude of quantum fluctuation of the curvaton field in a quasi-de Sitter space is given by
\begin{equation}
\delta\sigma=\frac{H_\ast}{2\pi},
\end{equation}
where $\ast$ denotes the epoch of horizon exit during inflation.
The curvature perturbation generated from curvaton is given by
\begin{equation}
P^{1/2}_{\zeta_\sigma}=\frac{1}{3}\Omega_{\sigma, D}\frac{H_\ast}{\sigma_\ast}
\label{eq1}
\end{equation}
where
\begin{equation}
\Omega_{\sigma, D} \equiv \left(\frac{\rho_\sigma}{\rho_{tot}}\right)_D
\end{equation}
is the density fraction of the curvaton density $\rho_\sigma$ relative to
the total density of the universe $\rho_{tot}$ at the time of curvaton
decay, denoted by $D$, and $\zeta_\sigma$ is the curvature perturbation of the curvaton field $\sigma$. The amount of non-Gaussianity is
characterized by the nonlinear parameter $f_{NL}$ given by
\begin{equation}
f_{NL}=\frac{5}{4 \Omega_{\sigma, D}}.
\label{eq3}
\end{equation}
This equation is valid only when $\Omega_{\sigma, D} \ll 1$.

If we assume that at the time $t_o$ of curvaton oscillation with
energy density $\rho_\sigma(t_o)=m^2\sigma^2_\ast/2$, the universe
is dominated by radiation (the decay products of inflaton) with
energy density $\rho_R(t_o)=3m^2 M_P^2$, where $M_P \equiv 2.4
\times 10^{18}$ GeV is the reduced Planck mass. Note that in order to have $\rho_\sigma(t_o) \ll \rho_R(t_o)$, we require $\sigma_\ast \ll M_P$. At the time of
curvaton decay $t_D$, the energy density of the universe is given by
$\rho_R(t_D)=3\Gamma^2M_P^2=\rho_R(t_o)(a(t_o)/a(t_D))^4$. Therefore
$a(t_D)/a(t_o)=(m/\Gamma)^{1/2}$ and $\Omega_{\sigma, D}$ is given
by
\begin{equation}
\Omega_{\sigma, D}=\frac{\rho_\sigma(t_o)}{\rho_R(t_o)}\frac{a(t_D)}{a(t_o)}=\frac{\sigma^2_\ast}{6 M_P^2} \left(\frac{m}{\Gamma}\right)^{1/2}=\frac{\sigma^2_\ast}{6 M_P^2} \frac{\sqrt{4 \pi}}{\lambda_\nu}
\label{eq4}
\end{equation}
where $m$ is the Right-Handed sneutrino mass and $\Gamma$ its decay rate.
If at the time of curvaton oscillation, the universe is dominated by oscillating inflaton field, we have
\begin{equation}
\Omega_{\sigma, D}=\frac{\sigma^2_\ast}{6 M_P^2} \left(\frac{\Gamma_d}{\Gamma}\right)^{1/2}=
\frac{\sigma^2_\ast}{6 M_P^2} \frac{\sqrt{4 \pi x}}{\lambda_\nu}=\frac{\sigma^2_\ast}{6 M_P^2} \frac{\sqrt{4 \pi}}{\lambda}
\label{eq5}
\end{equation}
where $\lambda \equiv \lambda_\nu/\sqrt{x}$ and $\Gamma_d \equiv xm$
is the inflaton decay rate which is smaller than $m$ with $x<1$. Notice that the form of Eqs. (\ref{eq4}) and (\ref{eq5}) are the same and $\lambda$ here is just a parameter used to describe the case when the inflaton decay rate is smaller than the curvaton mass.

\begin{figure}[htbp]
\begin{center}
\includegraphics[width=0.45\textwidth]{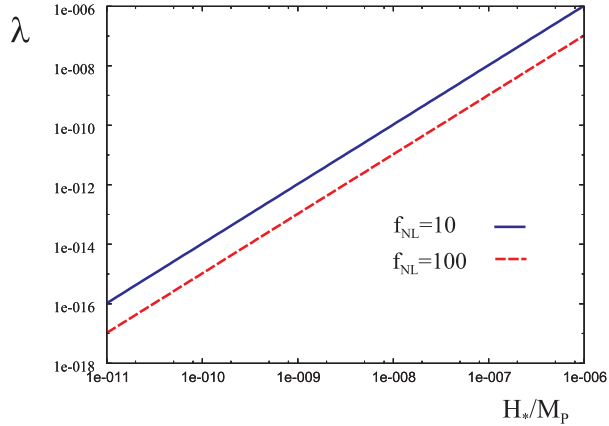}
\caption{$\lambda$ versus $H_\ast/M_P$}
\label{fig1}
\end{center}
\end{figure}

In this paper, we assume the curvature perturbation is dominated by
curvaton and the curvature perturbation from inflaton is negligible.
This liberates the constraint of inflation model building \cite{Dimopoulos:2002kt} and allows the
scale of inflation to be much lower. The condition is
\begin{equation}
P^{1/2}_{\zeta_{inf}}=\frac{1}{2\pi}\frac{H_\ast M_P}{\sqrt{\epsilon_H}} \ll 5 \times 10^{-5},
\end{equation}
where $\zeta_{inf}$ is the curvature perturbation generated from inflaton. For a typical value of $\epsilon_H \sim 0.01$, this implies $H_\ast \ll 10^{-5}M_P$. Using Eq. (\ref{eq1}) and Eq.
(\ref{eq5}) and imposing CMB normalization ($P^{1/2}_{\zeta_\sigma}
\simeq 5 \times 10^{-5}$), we obtain
\begin{equation}
\lambda=3.9 \times 10^3 \sigma_\ast H_\ast
\label{eq6}
\end{equation}
For conventional\footnote{If the curvaton is a Pseudo Nambu-Goldstone Boson (PNGB) with a symmetry-breaking phase transition during inflation or the curvaton mass increases suddenly at some moment after the end of inflation, it is possible to get $H_\ast$ as low as $1$ TeV \cite{Dimopoulos:2004yb, Dimopoulos:2005bx}.} curvaton scenario like our case,
there is a lower bound for the Hubble parameter during inflation,
$H_\ast \gae 10^7$ GeV $=4.17 \times 10^{-12}M_P$
\cite{Lyth:2003dt}. Hence we are interested in the range
\begin{equation}
10^{-11} \lae H_\ast/M_P \lae 10^{-6}.
\end{equation}
From Eq. (\ref{eq3}), Eq. (\ref{eq5}) and Eq. (\ref{eq6}), we have
\begin{equation}
f_{NL}=2.68 \times 10^3 \frac{H_\ast}{\sigma_\ast}
\label{eq8}
\end{equation}
In this paper, we are interested in large non-Gaussianity and will
explore the parameter space for $10 \lae f_{NL} \lae 100$. Using Eq.
(\ref{eq6}-\ref{eq8}), we can plot $\lambda$ versus $H_\ast/M_P$ as
shown in Fig. (\ref{fig1}).

Since curvaton must decay after inflaton $\Gamma=\lambda^2_\nu m/4\pi<\Gamma_d=xm$, we obtain $x>\lambda^2_\nu/4\pi$ and $\lambda=\lambda_\nu/\sqrt{x}<\sqrt{4\pi}$. A lower bound for $\Gamma$ is obtained from the requirement that curvaton should not disturb big bang nucleosynthesis (BBN), which introduces \cite{Dimopoulos:2003az}
\begin{equation}
\Gamma>4.5 \times 10^{-25}\mbox{GeV} =1.88\times 10^{-43}M_P.
\label{eq9}
\end{equation}
An upper bound for $\Gamma_d$ can be obtained from the gravitino bound
of reheating temperature if we assume the decay products of inflaton
are thermalized immediately after decay, then we have
\cite{McDonald:2003xq}
\begin{equation}
T_R=\sqrt{\frac{\Gamma_d M_P}{k_{T_R}}}<10^8\mbox{GeV}=4.17\times 10^{-11} M_P
\end{equation}
where $k_{T_R}=(4 \pi^{3} g(T_R)/45)^{1/2}$ and $g(T_R)$ is the effective number of massless degrees of freedom in thermal equilibrium. We will consider $k_{T_R} \approx 20$, corresponding to MSSM with $g(T_R) \approx 200$, this implies
\begin{equation}
\sqrt{\Gamma_d}=\sqrt{xm}<1.86 \times 10^{-10} M_P^{1/2}.
\label{eq11}
\end{equation}
Combining Eq. (\ref{eq9}) and (\ref{eq11}), we obtain the constraint
\begin{equation}
1.88\times 10^{-43}<\frac{\lambda_\nu^2 m}{4\pi}<xm<3.46\times 10^{-20}.
\label{eq19}
\end{equation}
From Eq. (\ref{eq11}) we can see that in order to evade the
gravitino bound, small $x$ is preferred, which means late decay of
the inflaton, however this will cause the Yukawa coupling
$\lambda_\nu$ to be suppressed from Fig. (\ref{fig1}). In Eq.
(\ref{eq19}), the upper bound can be relaxed if the inflaton does
not thermalized immediately after decay, or if there is other
methods to evade gravitino problem, for example, a period of thermal inflation \cite{Lyth:1995hj, Lyth:1995ka} after reheating..

\section{Initial condition}
\label{2}

There is no consensus in literature about what the most natural
initial condition for the curvaton field is. For example, if
non-renormalizable terms are not protected by any symmetry, the
slow roll condition may fail for large value of the curvaton field. This
provides an upper limit for the curvaton field value, then one may
choose the nature value of the curvaton field to be of the order of
its upper limit. Usually the picture is that in different patches of
the universes separated by the horizon, the curvaton field may take
on different values at horizon exit, therefore the curvature perturbation
and the amount of non-Gaussianities are different \cite{Linde:2005yw,
  Lyth:2006gd}.

In \cite{Dimopoulos:2003az, Dimopoulos:2003ss}, it is suggested that
the most likely value of the curvaton field may be determined by the
boundary between classical slow roll motion domination and quantum
fluctuation domination. In one Hubble time, the classical slow roll
gives a change $\Delta \sigma=-V^\prime/H^2_\ast$, while the quantum
fluctuation gives a random contribution $\Delta\sigma=\pm
H_\ast/2\pi$. When these two are equal, $|V^\prime|=H^3_\ast$ and so
gives $\sigma_\ast \sim H^3_\ast/m^2$.  This suggests a value for the
curvaton mass
\begin{equation}
m^2 \sim \frac{H^3_\ast}{\sigma_\ast}.
\end{equation}
We will refer this case as case 1.

\begin{figure}[htbp]
\begin{center}
\includegraphics[width=0.5\textwidth]{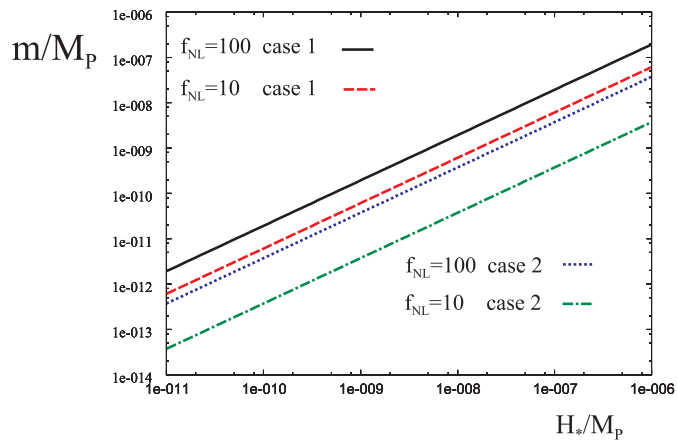}
\caption{$m/M_P$ versus $H_\ast/M_P$}
\label{fig2}
\end{center}
\end{figure}

Another different argument is given in \cite{Huang:2008ze} where the
author gave three different arguments to suggest that the typical
value of curvaton field is $\sigma_\ast \sim H^2_\ast/m$. This
suggests a curvaton mass
\begin{equation}
m \sim \frac{H^2_\ast}{\sigma_\ast}.
\end{equation}
We will refer this case as case 2.
We plot both cases in Fig. (\ref{fig2}).

We should emphasis that those are not strict constraints to the
allowed curvaton mass, however, the right-handed sneutrino mass may lie in the range where curvaton works best.

\section{The spectral index}
\label{3}
The spectral index $n_s$ in curvaton scenario takes on the form

\begin{equation}
n_s=1+2\eta_{\sigma\sigma}-2\epsilon,
\end{equation}
where
\begin{equation}
\eta_{\sigma\sigma} \equiv \frac{1}{3H^2}\frac{d^2 V(\sigma)}{d\sigma^2}  \;\;\;\;\mbox{and}\;\;\;\; \epsilon_H \equiv -\frac{\dot{H_\ast}}{H^2_\ast}.
\end{equation}
Here $\eta_{\sigma\sigma}$ should be evaluated at horizon exit. WMAP
data prefers a red tilted spectrum with spectral index $n_s \simeq
0.96$ \cite{Komatsu:2008hk}. Because a large positive
$\eta_{\sigma\sigma}$ will result in a blue spectrum, it is not
preferred. It is possible to get a
large negative $\eta_{\sigma\sigma}$ during inflation. This happens
for example from F-term hybrid inflation in which the vacuum energy
during inflation breaks supersymmetry and will introduce a soft mass
term of the order of the Hubble parameter. By choosing the magnitude
correctly\footnote{This may be justified if we allow the curvaton mass to run via the method of \cite{Stewart:1996ey, Stewart:1997wg}}, $n_s \simeq 0.96$ can be achieved \cite{McDonald:2003xq, Matsuda:2007av}.
We can imagine this also happens in D-term hybrid inflation if we
consider a non-minimal gauge kinetic function, because a similar
large mass correction can occur \cite{Lyth:1997ai, Lin:2008ys}.
Another possibility is that if we have $\epsilon \sim 0.02$, $n_s
\simeq 0.96$ can be achieved. However, this may not be easy to
achieve for some models, for example, \cite{Lin:2009yn}. In this
case, we will have $n_s \simeq 1$. In \cite{Parkinson:2006ku}, the
authors argued that we can put $n_s=1$ as the prior and show that it
is not ruled out by WMAP data. Another way out is to assume there
are some cosmic strings produced after inflaton as done in
\cite{Lin:2009yn}.

\section{Curvaton on the brane}
\label{4} As we can see from Eq.~(\ref{eq19}), it is generically
true that for a successful curvaton model, the Yukawa coupling is
very small. A very small Yukawa coupling may be regarded as
fine-tuning. In this section, we will show that if we consider our
world as a brane where gravity and matter field are confined on the
3-brane in the 5-dimensional spacetime \cite{Randall:1999vf}, the
model can work with a larger Yukawa coupling. In this set-up, the
Friedmann equation becomes \cite{Maartens:1999hf}
\begin{equation}
H^2 \equiv \left(\frac{\dot{a}}{a}\right)^2=\frac{\rho}{3M_P^2}\left(1+\frac{\rho}{2\Lambda}\right),
\end{equation}
where the brane tension $\Lambda$ relates the four dimensional Planck mass
($M_4=1.2\times 10^{19}\mbox{ GeV}$) to the five dimensional Planck
mass as $\Lambda=3M_5^6/4 \pi M^2_4$. The Big Bang Nucleosynthesis
(BBN) limit implies $M_5 \gae 10\mbox{ TeV} \sim 10^{-14}M_P $,
where $M_P=2.4\times 10^{18}\mbox{ GeV}$ is the reduced Planck mass.
If $\rho/2\Lambda>1$, we have $\rho\simeq \sqrt{6\Lambda}H M_P$, therefore $\rho_R(t_o)=\sqrt{6\Lambda}mM_P$ and $\rho_R(t_D)=\sqrt{6\Lambda}\Gamma M_P$. Repeat what we did in Sec.~\ref{1}, we found
\begin{equation}
\Omega_{\sigma,D}=\frac{m \sigma^2_\ast}{2\sqrt{6\Lambda}M_P}\left(\frac{m}{\Gamma}\right)^{1/4}=\frac{m \sigma^2_\ast}{2\sqrt{6\Lambda}M_P}\left(\frac{4\pi}{\lambda^2}\right)^{1/4}
\label{equation1}
\end{equation}
and
\begin{equation}
P_{\zeta_{\sigma}}^{1/2}=\frac{m \sigma_\ast}{6\sqrt{6\Lambda}M_P}\left(\frac{m}{\Gamma}\right)^{1/4}H_\ast=\frac{m \sigma_\ast}{6\sqrt{6\Lambda}M_P}\left(\frac{4\pi}{\lambda^2}\right)^{1/4}H_\ast.
\label{equation2}
\end{equation}
Here we have assumed $\rho/2\Lambda \lae 1$ before curvaton decay which implies
\begin{equation}
\Lambda \lae \frac{3}{2}\Gamma^2 M_P^2.
\end{equation}
For estimate, we saturate the inequality and by using Eqs. (\ref{equation1}) and (\ref{equation2}) we obtain
\begin{equation}
\lambda^{5/2}=2.63\times 10^{4}\frac{\sigma_\ast H_\ast}{M_P^2}.
\end{equation}
For comparison, if $f_{NL}=10$ (this implies $\sigma_\ast=8.3\times 10^2H_\ast$), hence
\begin{equation}
\lambda^{5/2}=6.58 \times 10^7H_\ast^2/M^2_P.
\end{equation}
Therefore, for example, when $H_\ast=10^{-8}M_P$, we obtain $\lambda=5.34\times 10^{-4}$, which is larger than the usual case.

\section{Conclusions}
\label{5}
In this paper we explored the allowed value of Yukawa coupling
$\lambda_\nu$ in the right-handed sneutrino curvaton scenario to
generate large non-Gaussianity ($10 \lae f_{NL} \lae 100$) within the
scale of inflation $10^{-11} \lae H_\ast \lae 10^{-6}$, which exhausted
the allowed range for a normal curvaton model. The Yukawa coupling
$\lambda_\nu$ will be further suppressed if the inflaton decay rate is
lower than curvaton mass which is favored if we want to evade the
gravitino problem in the framework of supersymmetry. We also consider
two different kinds of scenarios in literature considering the
most likely value of the curvaton field and show that these scenarios
suggest the mass of the RH sneutrino. The mass is not severely
constrained by this scenario, but it maybe of use if in the future we
can determine the RH neutrino mass in a different way. We have also showed
that in the case of braneworld scenario, the Yukawa coupling can be
larger which may make it a more natural model.

In this paper our arguments are focused on the right-handed sneutrino
curvaton, because it is a good candidate in the framework of
MSSM. However, our analysis can also apply to the case of any curvaton
models with a quadratic potential $V(\sigma)=m^2\sigma^2/2$ and decay
rate of the form $\Gamma=h^2m/4\pi$ where $h$ is some constant
corresponds to our $\lambda_\nu$. In the models where supersymmetry is
not imposed, there will be no gravitino bound for reheating
temperature and $h$ is less constrained.

\section*{Acknowledgement}
This work was supported in part by the
NSC under grant No. NSC 96-2628-M-007-002-MY3, by the NCTS, and by the
Boost Program of NTHU, and
by WCU program through the NRF funded by the MEST (R31-2008-000-10057-0).

\end{document}